\documentclass[prd,aps,a4paper,eqsecnum,twocolumn,nofootinbib,floatfix]{revtex4}  %

\newif\ifusesec
\usesectrue  
   
\usepackage{graphicx} 
\usepackage{amsmath,amsfonts,amssymb}
\usepackage[section]{placeins}
\usepackage{appendix}
\usepackage{mathtools}
\usepackage[caption=false,justification=justified]{subfig}
\usepackage{xcolor}
\usepackage{slashed}

\newcommand{\beq}{\begin{equation}}
\newcommand{\eeq}{\end{equation}}
\newcommand{\bea}{\begin{eqnarray}}
\newcommand{\eea}{\end{eqnarray}}
\newcommand{\nn}{\nonumber}

\begin{document}

\title{Radiative losses from unbound orbits at quadratic order in spin from MPM formalism}

\author{Donato Bini$^{1}$, Giorgio Di Russo$^{2}$}   
  \affiliation{
$^1$Istituto per le Applicazioni del Calcolo ``M. Picone,'' CNR, I-00185 Rome, Italy\\
$^2$School of Fundamental Physics and Mathematical Sciences, Hangzhou Institute for Advanced Study, UCAS, Hangzhou 310024, China\\
}

\date{\today}

\begin{abstract}
We study the spinning two-body system in the aligned spin case for hyperboliclike motions computing all radiative losses (at the 2PN absolute accuracy and including spin-squared corrections) using the MPM formalism and generalizing previous results valid at linear order in spin [Phys. Rev. D \textbf{108}, no.6, 064049 (2023)].
Leading PM order results are checked against existing literature, whereas higher-order PM results (within the 2PN accuracy) are new with this work.
As a by-product of our general results we analyze the spinning situation which supports radial fall (at 2PN and including spin-squared accuracy level contributions), showing that as soon as the PN accuracy increases deviations from radial fall appear necessarily.
\end{abstract}

\maketitle

\section{Introduction}
\label{Intro}

The ability of laser-interferometric gravitational-wave (GW) detectors, such as LIGO, Virgo, KAGRA, to extract meaningful astrophysical information from the  signals emitted by coalescing compact binary systems relies critically on the availability of highly accurate theoretical models describing their inspiral and merging phases.

The detection and analysis of the characteristic chirp waveform generated by these systems are based on matched-filtering techniques, which compare observational data with theoretical templates parameterized by the physical properties of the source. Tracking the theoretical evolution of the GW frequency as it sweeps through the detector's sensitive frequency band represents one of the most effective methods for determining the masses and spins of the binary components.

It is a matter of fact that all the GW signals detected up to now (which are on the order of 400 in number) correspond to coalescing spinning binaries, and therefore spin effects cannot be neglected in current theoretical studies \cite{LIGOScientific:2025rid}. For example, spin components that are not aligned with the orbital angular momentum generate orbital plane precession, leading to observable modulations of the gravitational waveform. Moreover, spin effects enter the radiative multipole moments, thereby influencing the waveform amplitude and phase evolution, as well as the fluxes of energy, angular momentum, and linear momentum emitted by the system through gravitational radiation.

In principle, besides the case of coalescing binaries one should also expect GW signals from scattering processes, and for this situation a growing body of research has been devoted to the development of increasingly accurate gravitational-wave templates. A large number of theoretical results is available for hyperbolic encounters in the non-spinning case, whereas only a limited number of studies have addressed the spinning case, primarily focusing on configurations in which the spins are aligned and perpendicular to the orbital plane, and therefore also aligned with the system's orbital angular momentum. 

Another interesting process which has received special attention in recent works (and which we will tackle also here including the spin of the bodies) is the radial infall. Although this problem has a long history, the vast majority of existing studies are either numerical \cite{Davis:1971gg,Detweiler:1979xr,Mitsou:2010jv} or semi-analytical \cite{Zerilli:1970se,Maggiore:2007ulw,Maggiore:2018sht}. Unfortunately, perturbative approaches allow one to follow an infalling trajectory only for a limited amount of time, say up to $t=t_{\rm max}$, since for $t>t_{\rm max}$ the particle inevitably enters the strong-field region, for instance as it approaches the black hole horizon. Recent analytical developments based on the Post-Newtonian (PN) approximation for the computation of the waveform emitted by radially infalling particles have been reported in \cite{Bini:2026ova,DiRusso:2026fqn}. However, the PN expansion is intrinsically limited to the weak field regime and therefore can describe only a portion of the trajectory. Efficient analytical tools capable of accurately treating the strong field regime, beyond numerical relativity, are still under development.

In the present study, we employ center-of-mass (cm) harmonic coordinates, where most results for the non-spinning case have been derived.
We limit to the aligned spin case but, preparing however all the building blocks to go beyond this simplified situation. Our results, aiming at computing all radiative losses, use Multipolar-Post-Minkowskian (MPM) formalism and are limited at the second PN approximation level (using the notation $\eta=\frac1{c}$ as a place-holder for PN expansion), which includes, however, spin-orbit (SO) and spin-spin (SS) contributions at the leading-order, but do not include hereditary (tail) effects and radiation-reaction effects yet. 
For a recent treatment of the conservative dynamics in a Hamiltonian context see for example \cite{Mandal:2022nty,Mandal:2022ufb}. Let us mention, in passing, that effects related to the spin-orbit coupling (because of the linearity in the spin) have received much attention in the literature, see e.g. \cite{Damour:2007nc,Damour:2008qf,Nagar:2011fx,Barausse:2011ys,Bini:2014ica,Bini:2015mza,Kavanagh:2015lva,Dolan:2013roa,Kavanagh:2017wot}. 

Alternative approaches to GW waveform modeling rely on the gravitational self force formalism \cite{Detweiler:2002mi,Barack:2018yvs}. Recently, this framework has been extended to regular compact objects, including a class of smooth solutions of five-dimensional Einstein--Maxwell theory known as Topological Stars \cite{Bianchi:2024vmi,Bianchi:2024rod,DiRusso:2025lip,Bianchi:2025aei}, as well as to W solitons \cite{Bianchi:2025ydq}. A remarkable property of these horizonless geometries is that, over a suitable range of parameters, they closely mimic black holes at large distances while remaining completely regular in their interior \cite{Bah:2020pdz,Dima:2025tjz}.

From a technical perspective, significant progress has also been made in the analytical treatment of the confluent Heun equation (CHE), which governs all linear perturbations of the Schwarzschild geometry. Recent developments include methods based on the instanton calculus of $\mathcal{N}=2$ $SU(2)$ super Yang--Mills (SYM) theory, commonly referred to as Seiberg--Witten (SW) theory \cite{Cipriani:2025ikx,Cipriani:2026xmx}, as well as techniques based on the analysis of the Floquet basis \cite{Fioravanti:2025bts}.

In this paper, we focus on hyperboliclike orbits obtaining by a direct integration of the equations of motion both spin-orbit (already known) and spin-squared (an original contribution of this work) modifications to the quasi-Keplerian representation of the motion. 

With the obtained orbit we check the conservation of energy and angular momentum. 
Then, using the MPM formalism at 2PN and $O(S^2)$ (leading-order in spin-spin corrections), we evaluate the various losses: energy, angular and linear momentum along spinning hyperboliclike orbits.
As a check we have re-obtained previous results from Effective Field Theory (EFT) \footnote{EFT has obtained the most updated PN expanded results, reaching also the fourth order in spin accuracy but has mainly considered applications to elliptic-like motions, see e.g., \cite{Cho:2022syn,Cho:2021mqw,Liu:2021zxr}. MPM formalism (the one used here) has not developed yet the analogous information (out of special situations like circular orbits) for a complete comparison.} and amplitudes, i.e., the leading order PM result valid at all PN orders \cite{Jakobsen:2021lvp,Riva:2022fru}. Viceversa, working at the leading PN order we computed here all the various PM order contributions preparing valuable materials to support future accomplishments.

Finally, as a further  application (as already mentioned), we have considered the radial infall process, which, for what concerns the orbit, takes leading corrections at order spin squared, and not at the linear-in-spin level. In this case, we have obtained (still at 2PN and $O(S^2)$) the orbit and the radiative losses in PN sense, i.e. in the region of validity of the PN approximation (weak field and slow motions). Indeed,  our results  give only a partial (PN) answer to the problem, complicated by the fact that as approaching the horizon the gravitational field becomes strong and one necessarily exits the PN regime. As it is well known, out of the PN approximation, this kind of computations are still challenging and no analytic results exist yet (while a number of numerical results are already available). Furthermore, the presence of the spin, as arguable, imply deviations from radial infall as soon as the PN description of the orbit increases.

\section{Setting of the problem}

Let us denote the masses of the bodies as $m_1$ and $m_2$ (assuming $m_1>m_2$, i.e., not considering the equal mass case) 
 and the spins as ${\mathbf S}_1$ and ${\mathbf S}_2$.
Standard notations for the masses are
\beq
M\equiv m_1+m_2\,,\qquad\delta m \equiv m_1 - m_2, 
\eeq
so that
\beq
m_1=\frac{M}{2}(1+\sqrt{1-4\nu})\,,\quad m_2=\frac{M}{2}(1-\sqrt{1-4\nu})\,,
\eeq
with $\delta m=M\sqrt{1-4\nu}$. We will also introduce the dimensionless ratios
\beq
X_1=\frac{m_1}{M}\,,\qquad X_2=\frac{m_2}{M}\,,
\eeq
implying
\beq
X_1+X_2=1\,,\qquad \frac{\delta m }{M}=X_1-X_2\,.
\eeq

Standard notations for the spins involve the following $1-2$ symmetric combinations 
\beq\label{ssstar}
{\mathbf S}={\mathbf S}_1+{\mathbf S}_2\,,\qquad {\mathbf S}_*=\frac{m_2}{m_1}{\mathbf S}_1+\frac{m_1}{m_2}{\mathbf S}_2\,,
\eeq
with their dimensionless counterparts given by
\beq
\hat {\mathbf S}=\frac{{\mathbf S}}{M^2}\,,\qquad \hat {\mathbf S}_*=\frac{{\mathbf S}_*}{M^2}\,.
\eeq
Consequently, for example,
\bea
{\mathbf S}+{\mathbf S}_*&=& M\left(\frac{{\mathbf S}_1}{m_1}+ \frac{{\mathbf S}_2}{m_2}\right)\,,\nonumber\\
{\mathbf S}-{\mathbf S}_*&=& \delta m \left(\frac{{\mathbf S}_1}{m_1}-\frac{{\mathbf S}_2}{m_2} \right)\,.
\eea

Inverting Eq. \eqref{ssstar} leads to
\bea
\mathbf{S}_1
&=& \frac{X_1}{X_1-X_2}\left(X_1 \mathbf{S} -X_2 \mathbf{S}_* \right)\,,\nonumber\\
\mathbf{S}_2
&=&  \frac{X_2}{X_1-X_2}\left(-X_2 \mathbf{S} +X_1 \mathbf{S}_* \right)
\,.
\eea
In the literature one also finds the following  combination
\beq
{\bf \Delta} \equiv M \left(\frac{{\bf S_2}}{m_2} -\frac{{\bf S_1}}{m_1}\right)= \frac{M}{\delta m} (\mathbf{S}_* -\mathbf{S})  \,,
\eeq
such that
\beq
\frac{\delta m}{M} {\bf \Delta}=\mathbf{S}_* -\mathbf{S}\,.
\eeq
In addition, one often defines the spins per unit of corresponding mass
\beq
{\boldsymbol {\sf a}}_1=\frac{{\mathbf S}_1}{m_1}\,,\quad {\boldsymbol {\sf a}}_2=\frac{{\mathbf S}_2}{m_2}\,.
\eeq
However, these symbols ${\boldsymbol {\sf a}}_i$ can be confused with the accelerations and hence will be avoided hereafter.

The equations of motion for two spinning bodies have been
developed by numerous authors. 
By eliminating the cm of the system (denoting by ${\bf x} \equiv {\bf y_1}-{\bf y_2}$ the relative position, ${\bf v}={\bf v}_1-{\bf v}_2={d{\bf x}/dt}$ the relative velocity,
${\mathbf n}\equiv{{\bf x}/r}\equiv {\mathbf n}_{12}$ (with $r=r_{12}=|{\bf x}|$); 
as standard, an overdot
denotes differentiation with respect to $t$), one converts the two body
equations of motion to a relative one-body equation of motion of reduced mass $\mu=m_1m_2/M=M\nu$. The cm relative acceleration, ${\mathbf a}={\mathbf a}_1-{\mathbf a}_2$, using harmonic coordinates and truncating at the 2.5PN level of accuracy, reads \cite{Gergely:1999pd}
\begin{equation}
\label{rel_acc}
{\mathbf a} = {\mathbf a}_{\rm N} + {\mathbf a}_{\rm 1PN} +\underbrace{ {\mathbf a}_{\rm SO}}_{\rm 1.5PN} +  \underbrace{{\mathbf a}_{\rm 2PN}
+{\mathbf a}_{\rm SS}}_{\rm 2PN} + \underbrace{{\mathbf a}_{\rm rr}+{\mathbf a}_{\rm SO}^{\rm NLO}}_{\rm 2.5PN}+\ldots\,, 
\end{equation}
where
\begin{widetext}
\bea\label{acomp}
{\bf a}_{\rm N} &=& - \frac{M}{r^2} {\mathbf n}\,, \nonumber\\
{\bf a}_{\rm 1PN} &=&  - \frac{M}{r^2} \left(  {\mathcal A}^{\rm 1PN}{\mathbf n} 
+{\mathcal B}^{\rm 1PN}\dot r {\bf v} \right)\,,\nonumber\\
{\bf a}_{\rm SO} &=& \frac{1}{r^3} \left[ 6 {\mathbf n}\,  ( {\mathbf n} \times
{\bf v} ) {\bf \cdot} ({\bf S} + {\mathbf S}_*) 
-  {\bf v} \times \Big(4
{\bf S}+3{\mathbf S}_*\Big) 
+ 3 \dot r   {\mathbf n} \times
\Big(2{\bf S} + {\mathbf S}_* \Big)  \right]\,,\nonumber\\
{\bf a}_{\rm 2PN} &=& - \frac{M}{r^2} \biggl({\mathcal A}^{\rm 2PN}{\mathbf n} 
+{\mathcal B}^{\rm 2PN}\dot r {\bf v} \biggr)\,,\nonumber\\
{\bf a}_{\rm SS} &=& - \frac{3}{\mu r^4} \biggl[ {\mathbf n} ({\bf S_1 \cdot
S_2}) + {\bf S_1} ({\bf  n \cdot S_2}) + {\bf S_2} ({\bf n \cdot
S_1}) - 5 {\mathbf n} ({\bf  n \cdot S_1})({\bf n \cdot S_2})
\biggr]\nonumber\\
&-& \frac{3}{2 \mu r^4} \left[ {\mathbf n} \left(\frac{m_2}{m_1} S_1^2 + \frac{m_1}{m_2} S_2^2\right) + 
   2 \left(\frac{m_2}{m_1} ({\mathbf n} \cdot {\bf S_1}) {\bf S_1} + \frac{m_1}{m_2} ({\mathbf n} \cdot {\bf S_2}) {\bf S_2}\right)\right.\nonumber\\ 
&-&\left.  
   5 {\mathbf n} \left( \frac{m_2}{m_1} ( {\mathbf n}\cdot  {\bf S_1})^2 +  \frac{m_1}{m_2} ({\mathbf n}\cdot {\bf S_2})^2\right)\right]
\,,
\eea
\end{widetext}
with
\bea
{\mathcal A}^{\rm 1PN}&=& 
(1+3\nu)v^2 - 2(2+\nu)\frac{M}{r} - \frac{3}{2} \nu \dot r^2\,,  \nonumber\\
{\mathcal B}^{\rm 1PN}&=& -2(2-\nu)\,, \nonumber\\
{\mathcal A}^{\rm 2PN}&=& \frac{3}{4}
(12+29\nu) \left( \frac{M}{r} \right)^2
+ \nu(3-4\nu)v^4 \nonumber\\
&+& \frac{15}{8} \nu(1-3\nu)\dot r^4- \frac{3}{2} \nu(3-4\nu)v^2 \dot r^2
\nonumber \\ 
&- &
\frac{1}{2} \nu(13-4\nu) \frac{M}{r} v^2
- (2+25\nu+2\nu^2) \frac{M}{r} \dot r^2 \,,\nonumber\\
{\mathcal B}^{\rm 2PN}&=& - \frac{1}{2} 
\left[ \nu(15+4\nu)v^2 - (4+41\nu+8\nu^2)
\frac{M}{r}\right.\nonumber\\
&-& \left. 3\nu(3+2\nu) \dot r^2 \right]\,,
\eea
where ${\bf a}_{\rm N}$, ${\bf a}_{\rm 1PN}$, and ${\bf a}_{\rm 2PN}$ are the Newtonian,
1PN, and 2PN contributions to the
equations of motion, and
${\bf a}_{\rm SO}$ and ${\bf a}_{\rm SS}$ are the spin-orbit and spin-spin
contributions to the equations of motion \cite{Kidder:1995zr}.  
Going beyond the 2PN approximation level requires considering additional contributions. For example,  at 2.5PN, one should include a LO radiation-reaction relative acceleration  
\bea
{\bf a}_{\rm rr} &=& \frac{8}{5} \nu \frac{M^2}{r^3} \left[-\mathbf{v}\Big(v^2+\frac{3M}{r}\Big)+\dot r \mathbf{n}\Big(3v^2+\frac{17M}{3r}\Big) \right]\,,\nonumber\\
\eea
as well as the NLO   in the spin-orbit coupling \cite{Faye:2006gx},
\begin{widetext}
\bea
{\bf a}_{\rm SO}^{\rm NLO}&=&\frac{1}{r^3}\Bigg\{{\bf n}\Big[\,{\bf S}\cdot({\bf n}\times {\bf v})\Big(-30\nu(nv)^2+24\nu v^2-\frac{M}{r}(38+25\nu)\Big)\nn\\
&+&\frac{\delta m}{M}\Delta\cdot({\bf n}\times{\bf v})\Big(-15\nu(nv)^2+12\nu v^2-\frac{M}{r}\Big(18+\frac{29}{2}\nu\Big)\Big)\big] \nn\\
&+&(nv){\bf v}\Big[{\bf S}\cdot({\bf n}\times {\bf v})(9\nu-9)+\frac{\delta m}{M}{\bf \Delta}\cdot({\bf n}\times{\bf v})(6\nu-3)\Big]\nn\\
&+&{\bf n}\times{\bf v}\Big[(nv)(vS)(3\nu-3)-\frac{8M}{r}\nu(nS)-\frac{\delta m}{M}\Big(\frac{4M}{r}\nu(n\Delta)+3(nv)(v\Delta)\Big)\Big]\nn\\
&+&(nv){\bf n}\times{\bf S}\Big[-\frac{45}{2}\nu(nv)^2+21\nu v^2-\frac{M}{r}(25+15\nu)\Big]\nn\\
&+&\frac{\delta m}{M}(nv){\bf n}\times{\bf \Delta}\Big[-15\nu(nv)^2+12\nu v^2-\frac{M}{r}\Big(9+\frac{17}{2}\nu\Big)\Big]\nn\\
&+&{\bf v}\times {\bf S}\Big[\frac{33}{2}\nu(nv)^2+\frac{M}{r}(21+9\nu)-14\nu v^2\Big]\nn\\
&+&\frac{\delta m}{M}{\bf v}\times{\bf \Delta}\Big[9\nu(nv)^2-7\nu v^2+\frac{M}{r}\Big(9+\frac{9}{2}\nu\Big)\Big]\Bigg\}\,.
\eea
\end{widetext}

In the present work we will stay at 2PN, in the sense that  we will evaluate the conservative dynamics at (absolute) 2PN level and all losses at (fractional) 2PN level, including however linear and quadratic in spin contributions. Henceforth, we will not consider LO radiation-reaction  nor NLO contributions to the spin-orbit (nor any type of hereditary effects), which will be, instead, the topic of future studies.

The main accomplishments of this paper are exact expressions of the radiative losses at order spin-square (in the aligned spin case and along hyperboliclike orbits) which have not been computed before.

The following points are worth noting.

The above expression for ${\bf a}_{\rm SO}$ is not
unique; it depends on a ``spin supplementary condition" (SSC) which is related
to the definition of the cm world line $x_A^\mu$ for each body $A$.
The above form of ${\bf a}_{\rm SO}$ is for the covariant (Pirani) SSC given by
$S_A^{\mu\nu} {u_A}_{\nu} = 0$, where
$u_A^\mu$ is the four-velocity of the cm world line of body $A$,
and
\begin{equation}
S_A^{\mu\nu} \equiv 2 \int_A (x^{[\mu} - {x_A}^{[\mu}) \tau^{\nu ] 0}
 d^3x , \label{spintensor}
\end{equation}
where $\tau^{\mu\nu}$ denotes the stress-energy tensor of matter plus
gravitational fields satisfying ${\tau^{\mu\nu}}_{,\nu}=0$, and
square brackets around indices denote antisymmetrization.
Moreover,  the spin vector $\bf S_A$
of each body is defined by
$S_A^i = \frac{1}{2} \epsilon_{ijk}S_A^{jk}$.
Let us emphasize that since we have chosen a cm world line
for each body through our choice of a SSC, we must ensure that all our
calculations are consistent with this choice.

It is interesting to note that while ${\bf a}_{\rm N}$, ${\bf a}_{\rm 1PN}$,
${\bf a}_{\rm 2PN}$, and ${\bf a}_{\rm rr}$ are all confined to the orbital plane,
in general ${\bf a}_{\rm SO}$ and ${\bf a}_{\rm SS}$ are not.  As a result, the
orbital plane will precess in space (except for specific spin orientations)
resulting in modulations of the observed waveform.

\section{Conserved energy and total angular momentum at 2PN}

In absence of radiation-reaction, the equations of motion admit a conserved energy and a conserved total angular momentum ${\mathbf J}={\mathbf L}+{\mathbf S}$ (see, e.g.,  Ref. \cite{Kidder:1992fr,Kidder:1993zz,Kidder:1995zr}. Note that the latter references are incomplete at order $O(S^2)$ since $O(S_{1,2}^2)$ are omitted)
\bea
E&=& E_{\rm N}+\eta^2 E_{\rm 1PN}+\eta^3 E_{\rm SO}+\eta^4 (E_{\rm 2PN}+E_{\rm SS})\,,\nonumber\\
{\mathbf L}&=& {\mathbf L}_{\rm N}+\eta^2 {\mathbf L}_{\rm 1PN}+\eta^3 {\mathbf L}_{\rm SO}+\eta^4 {\mathbf L}_{\rm 2PN}\,,
\eea
where $E_{\rm N}$, $E_{\rm 1PN}$, $E_{\rm 2PN}$,  ${\mathbf L}_{\rm N}$, ${\mathbf L}_{\rm 1PN}$, and ${\mathbf L}_{\rm 2PN}$ can be found in various references (see, e.g., Ref. \cite{Blanchet:2013haa}), while
\bea
E_{\rm SO}&=& \frac{G}{c^2 r^3}{\mathbf L}_{\rm N}\cdot {\mathbf S}_*\,,\nonumber\\
E_{\rm SS}&=& \frac{G}{c^2 r^3} [3(nS_1)(nS_2)-(S_1S_2)]\nn\\
&-&\frac{1}{2r^3}\Big(\frac{m_2}{m_1}S_1^2+\frac{m_1}{m_2}S_2^2\Big)\nonumber\\ 
&=& \frac{G \nu}{c^2 r^3 (1-4\nu)}\left[ -3\nu  (nS)^2 - 3\nu  (nS_*)^2 \right.\nonumber\\ 
&+& 3 (nS) (nS_*) (1 - 2 \nu) + 
 \nu S^2 + \nu S_*^2 \nonumber\\
&-&\left. (1 -2 \nu) (S S_*)\right]\nn\\
&-&\frac{G \nu}{2c^2 r^3(1-4\nu)}\Big[S^2+S_*^2-2\nu(S+S_*)^2\Big]\,,\nonumber\\
{\mathbf L}_{\rm SO}&=&\frac{\mu}{M}\frac{G}{c^2}\left\{\frac{M}{r}{\mathbf n}\times \left[{\mathbf n}\times(2{\mathbf S}+{\mathbf S}_*)\right]\right.\nonumber\\ 
&-&\left. \frac12 {\mathbf v}\times [{\mathbf v}\times {\mathbf S}_*] \right\} \,,
\eea
and there is no ${\mathbf L}_{\rm SS}$ at the 2PN order.
We have checked that both energy and total angular momentum ${\mathbf J}$  of the system 
are constant along the motion.

\section{Radiated energy, angular momentum and linear momentum at fractional 2PN from MPM}

We summarize in Tables below the status of the art concerning spin corrections  to the various source multipole moments up to spin-square terms (at leading-order) included. We use as spin variables the combinations ${\mathbf S}$ and ${\mathbf S}_*$.
The spinless part of the source multipole moments in the cm and written using harmonic coordinates are omitted, since they are now available from many review papers, including the LRR \cite{Blanchet:2013haa}.
\begin{table*}
\caption{\label{Spin_corr_multipoles} We take the expressions of $I_{ij}^{S}$ and $J_{ij}^{S}$ from Ref.~\cite{Blanchet:2006gy}, Eqs.~(5.3) and (5.4), respectively, while those for $I_{ij}^{SS}$ and $J_{ij}^{SS}$ are taken from Ref.~\cite{Bohe:2015ana}, Eqs.~(B.1) and (B.2), respectively. All  of them are rewritten in terms of our spin variables  $S$  and $S_*$ used in the present paper. We omit the parts of the various multipoles not containing the spin (or even the complete multipoles when the spin corrections are not needed: these expressions can be found e.g. in Ref. \cite{Blanchet:2013haa,Faye:2012we}). Cubic order in spin contribution are available in \cite{Marsat:2014xea}. Here $(ab)$ denotes the scalar product of ${\mathbf a}$ and ${\mathbf b}$.}
\begin{ruledtabular}
\begin{tabular}{ll}
$I_{ij}^{S}$ & $\nu\eta^3r\Bigg\{\frac{8}{3}n^{\langle i}({\bf v}\times{\bf S}_*)^{j\rangle}-\frac{4}{3}v^{\langle i}({\bf n}\times{\bf S}_*)^{j\rangle}\Bigg\}$\\
&$+ \nu\eta^5\Bigg\{(  
\frac{5}{3}+\frac{2}{7}\nu)  
M(nv)n^{\langle i}({\bf n}\times{\bf S}_*)^{j \rangle}+(-\frac{4}{21}+\frac{13}{21}\nu)M(nv)n^{\langle i}({\bf n}\times{\bf S})^{j\rangle }$\\
&$+ \Bigg((\frac{7}{3}+4\nu)\frac{M}{r}+(\frac{26}{21}-\frac{116}{21}\nu)v^2\Bigg)rn^{\langle i}({\bf v}\times{\bf S}_*)^{j\rangle}+\Bigg((\frac{26}{21}-\frac{28\nu}{21})\frac{M}{r}+\frac{38\nu}{21}v^2\Bigg)rn^{\langle i}({\bf v}\times {\bf S})^{j\rangle}$\\
&$+\Bigg((-4-\frac{2}{7}\nu)\frac{M}{r}+(-\frac{6}{7}+\frac{64}{21}\nu)v^2\Bigg)rv^{\langle i}({\bf n}\times {\bf S}_*)^{j\rangle}\Bigg((-\frac{14}{3}-\frac{8\nu}{21})\frac{M}{r}-\frac{10\nu}{21}v^2\Bigg)rv^{\langle i}({\bf n}\times {\bf S})^{j\rangle}$\\
&$+ (\frac{10}{21}-\frac{8\nu}{21})r(nv)v^{\langle i}({\bf v}\times{\bf S}_*)^{j\rangle}-\frac{22\nu}{21}r(nv)v^{\langle i}({\bf v}\times {\bf S})^{j\rangle}$\\
&${+}\Bigg((\frac{62}{21}{-}\frac{18\nu}{7})(S_*,n,v){+}(-\frac{10}{21}{+}\frac{8\nu}{7})(S,n,v)\Bigg)M n^{\langle i}n^{j\rangle}{+}\Bigg(({-}\frac{5}{21}{-}\frac{4\nu}{7})(S_*,n,v){+}\frac{9\nu}{7}(S,n,v)\Bigg)r v^{\langle i}v^{j\rangle}$\\
&${+} \Bigg(({-}\frac{8}{3}{+}\frac{8\nu}{3})(nS_*){+}\frac{8\nu}{3}(nS)\Bigg)Mn^{\langle i}({\bf n}\times {\bf v})^{j\rangle}{+}\Bigg((\frac{4}{3}{-}\frac{8\nu}{3})(vS_*){-}\frac{4\nu}{3}(vS)\Bigg)rv^{\langle i}({\bf n}\times{\bf v})^{j\rangle}\Bigg\}{+}O\left(\eta^7\right)$\\
\hline
$I_{ij}^{SS}$ & $-\frac{\eta^4}{M}\Bigg[\Bigg(1+\nu\Bigg(\frac{M}{\delta m}\Bigg)^2\Bigg)S^{\langle i}S^{j\rangle}+\nu\left(\frac{M}{\delta m}\right)^2S_*^{\langle i}S_*^{j\rangle}-2\nu\left(\frac{M}{\delta m}\right)^2S_*^{\langle i}S^{j\rangle}\Bigg] +O\left(\eta^6\right)$\\
\hline
$J_{ij}^{S}$&$ \nu \frac{M}{\delta m}\eta \Bigg\{-\frac{3}{2}rn^{\langle i}S_*^{j\rangle}+\frac{3}{2}rn^{\langle i}S^{j\rangle}\Bigg\}+\nu\eta^3\Bigg\{(\frac{3}{7}-\frac{16\nu}{7})\frac{M}{\delta m}r(nv)v^{\langle i}S_*^{j\rangle}-(\frac{3}{7}-\frac{16\nu}{7})\frac{M}{\delta m}r(nv)v^{\langle i}S^{j\rangle}$\\
&$+\frac{3}{7}\frac{\delta m}{M}r(nv)v^{\langle i}S^{j\rangle}+\Bigg((\frac{27}{14}-\frac{109}{14}\nu)\frac{M}{\delta m}(vS_*)-(\frac{27}{14}-\frac{109}{14}\nu)\frac{M}{\delta m}(vS)+\frac{27}{14}\frac{\delta m}{M}(vS)\Bigg)r n^{\langle i}v^{j\rangle}$\\
&$+\Bigg((-\frac{11}{14}+\frac{47}{14}\nu)\frac{M}{\delta m}r(nS_*)-(-\frac{11}{14}+\frac{47}{14}\nu)\Bigg]\frac{M}{\delta m}r(nS)-\frac{11}{14}\frac{\delta m}{M}r(nS)\Bigg)v^{\langle i}v^{j \rangle}$\\
&$+\Bigg((\frac{19}{28}+\frac{13\nu}{28})\frac{M}{r}+(-\frac{29}{28}+\frac{143}{28}\nu)v^2\Bigg)\frac{M}{\delta m}rn^{\langle i}(S_*-S)^{j\rangle}$\\
&$+\Bigg((-\frac{4}{7}+\frac{31}{14}\nu)\frac{M}{\delta m}(nS_*)-(-\frac{4}{7}+\frac{31}{14}\nu)\frac{M}{\delta m}(nS)-\frac{29}{14}\frac{\delta m}{M}(n S)\Bigg)M n^{\langle i}n^{j\rangle}$\\
&$+ \Bigg(-\frac{1}{14}\frac{M}{r}-\frac{2}{7}v^2\Bigg)\frac{\delta m}{M}r n^{\langle i}S^{j\rangle}\Bigg\}+O\left(\eta^5\right)\,,$\\ 
\hline
$J_{ij}^{SS}$&$ \frac{\nu\eta^4}{2\delta m}\Bigg[-3S^{\langle i}({\bf S}_*\times {\bf v})^{j\rangle}+S^{\langle i}({\bf S}\times{\bf v})^{j\rangle}+2S_*^{\langle i}({\bf S}_*\times {\bf v})^{j\rangle}\Bigg]+O\left(\eta^5\right)$\\
\end{tabular}
\end{ruledtabular}
\end{table*}

\begin{table*}
\caption{\label{Spin_corr_multipoles2}  
The expressions used here are taken from the literature and re-expressed in terms of our spin variables $S$ and $S_*$. Specifically, $I_{ijk}^{S}$ and $J_{ijk}^{S}$ are obtained from Ref.~\cite{Blanchet:2006gy}, Eqs.~(5.5a) and (5.5b), respectively; $I_{ijk}^{SS}$ is taken from Ref.~\cite{Bohe:2015ana}, Eq.~(B.3); $I_{ijkl}^{S}$ from Ref.~\cite{Bohe:2013cla}, Eq.~(3.10e); and $I_{ijkl}^{SS}$ and $J_{ijk}^{SS}$ from Ref.~\cite{Henry:2022dzx}, Eqs.~(3.19e) and (3.19f), respectively.
}
\begin{ruledtabular}
\begin{tabular}{ll}
$I_{ijk}^{S}$& $\nu\eta^3r^2\Bigg\{\Bigg[-\frac{9}{2}\frac{\delta m}{M}+\frac{3}{2}(3-11\nu)\frac{M}{\delta m}\Bigg]n^{\langle i}n^j({\bf v}\times{\bf S})^{k\rangle }+3\Bigg[\frac{\delta m}{M}-(1-3\nu)\frac{M}{\delta m}\Bigg]n^{\langle i}v^j({\bf n}\times{\bf S})^{k\rangle}$\\
&$-\frac{3}{2}(3-11\nu)\frac{M}{\delta m}n^{\langle i}n^j({\bf v}\times{\bf S}_*)^{k\rangle}+3(1-3\nu)\frac{M}{\delta m}n^{\langle i}v^j({\bf n}\times{\bf S}_*)^{k\rangle}\Bigg\}+O\left(\eta^5\right)$\\
\hline
$I_{ijk}^{SS}$&$ \frac{3\nu r}{\delta m} \eta^4\Bigg(-n^{\langle i}S^jS^{k\rangle}+n^{\langle i}S_*^jS_*^{k\rangle}\Bigg)+O\left(\eta^5\right)$\\
\hline
$I_{ijkl}^{S}$ &$\nu\eta^3r^3\Bigg\{\frac{12\nu}{5}({\bf S}\times{\bf v})^{\langle i}n^jn^kn^{l\rangle}+({\bf S_*}\times{\bf v})^{\langle i}n^jn^kn^{l\rangle}\Bigg(-\frac{32}{5}+\frac{84}{5}\nu\Bigg)+\frac{24\nu}{5}({\bf n}\times{\bf S})^{\langle i}n^j n^k v^{l\rangle}$\\
&$+ ({\bf n}\times{\bf S}_*)^{\langle i}n^jn^kv^{l\rangle}\Bigg(-\frac{24}{5}+\frac{48\nu}{5}\Bigg)\Bigg\}+O\left(\eta^5\right)$\\
\hline
$I_{ijkl}^{SS}$ & $-\frac{6\nu r^2}{M}\eta^4\Bigg\{\Bigg[-1+\left(\frac{M}{\delta m}\right)^2(1-3\nu)\Bigg]S^{\langle i}S^jn^kn^{l\rangle}+2\Bigg[1-\left(\frac{M}{\delta m}\right)^2(1-3\nu)\Bigg]S^{\langle i}S_*^jn^kn^{l\rangle}$\\
&$+ \left(\frac{M}{\delta m}\right)^2(1-3\nu)S_*^{\langle i}S_*^jn^kn^{l\rangle}\Bigg\}+O\left(\eta^5\right)$\\
\hline
$J_{ijk}^{S}$&$2\nu\eta r^2n^{\langle i}n^kS_*^{k\rangle}+O\left(\eta^3\right)$\\
\hline
$J_{ijk}^{SS}$&$\frac{2\nu r}{M}\eta^4\Bigg\{\Bigg[1-\left(\frac{M}{\delta m}\right)^2(1-3\nu)\Bigg]({\bf v}\times {\bf S})^{\langle i}S_*^jn^{k\rangle}+\Bigg[1-\left(\frac{M}{\delta m}\right)^2(1-3\nu)\Bigg]({\bf v}\times{\bf S}_*)^{\langle i}S^jn^{k\rangle}$\\
&$-\Bigg[1-\left(\frac{M}{\delta m}\right)^2(1-3\nu)\Bigg]({\bf v}\times {\bf S})^{\langle i}S^jn^{k\rangle}+\left(\frac{M}{\delta m}\right)^2(1-3\nu)({\bf v}\times{\bf S}_*)^{\langle i}S_*^jn^{k\rangle}\Bigg\}+O\left(\eta^6\right)$\\
\end{tabular}
\end{ruledtabular}
\end{table*}

Finally, let us recall the relation between the angular momentum and the impact parameter $b$ in presence of  spin \footnote{A typo has affected Eq. (24) of Ref. \cite{Bini:2023mdz}, whereas the correct expression is rewritten here in Eq. \eqref{relLb}.} 
\bea\label{relLb}
L &=& \frac{b\sqrt{\gamma^2-1}}{Mh} +\frac{(E_1-m_1) {\sf a}_1+(E_2-m_2) {\sf a}_2}{m_1m_2}\nonumber\\
&=&  \frac{b\sqrt{\gamma^2-1}}{Mh} +\frac{h-1}{2m_1m_2 h}[(h-1)S+(h+1)S_*]\,,\nonumber\\
\eea
with $L=c{\mathcal L}/(Gm_1m_2)$ dimensionless, $E=Mh$, $h=\sqrt{1+2\nu(\gamma-1)}$, $\gamma=\sqrt{1+p_\infty^2}$, ${\sf a}_1=\frac{S_1}{m_1}$, ${\sf a}_2=\frac{S_2}{m_2}$ and
\bea
\frac{E_1}{m_1}=\frac{m_1+m_2\gamma}{E}\,,\qquad \frac{E_2}{m_2}=\frac{m_2+m_1\gamma}{E}\,.
\eea
Notice that, as explained in \cite{Vines:2017hyw}, Eq. \eqref{relLb} results from the change of worldline associated with the \lq\lq covariant" and \lq\lq canonical" spin supplementary conditions. This transformation is linear in the spin (see Eq. (104) of \cite{Vines:2017hyw}). 
Consequently, Eq. \eqref{relLb} does not receive quadratic-in-spin corrections.

\section{Hyperbolic-like orbit in the aligned spin case}

For hyperbolic like motion it is convenient to use the quasi-Keplerian representation \cite{Damour:1981bh,dd,DD1981a,Memmesheimer:2004cv,Cho:2018upo}, i.e.
\bea\label{keplerpar}
l=\bar{n}t&=&e_t \sinh u-u+g_{t}(V-u)+f_t\sin V\,,\nonumber\\
r&=&\bar{a}_r(e_r\cosh u-1)\,,\nonumber\\
\phi&=&K(V+f_\phi \sin(2V)+g_\phi \sin(3V))\,,
\eea
here truncated at the 2PN level included, and where
\bea
V&=&2\arctan\left(\sqrt{\frac{e_r+1}{e_r-1}}\tanh\left(\frac{u}{2}\right)\right)\,.
\eea

In the present conservative and spin aligned case we assume
\bea
{\mathbf S}=(0,0,S)\,,\qquad {\mathbf S}_*=(0,0,S_*)\,.
\eea
The orbital parameters will depend on spin. For example, 
\bea
\label{deco_ar_er}
a_r&=& a_r^{\rm orb}+\epsilon a_r^{S}+\epsilon^2 a_r^{SS}\,,\nonumber\\
e_r&=& e_r^{\rm orb}+\epsilon e_r^{S}+\epsilon^2 e_r^{SS}\,.
\eea
A complete list of these parameters is given in Table \ref{orbital_param} below. 
%
%
\begin{table*}
\caption{\label{orbital_param}  Orbital parameters in the aligned spin case including spin squared corrections. 
}
\begin{ruledtabular}
\begin{tabular}{ll}
$e_\phi$&$e_r{-}\frac{e_r\nu}{2a_r}\eta^2{+}\frac{e_rS_*}{a_r^\frac{3}{2}\sqrt{e_r^2{-}1}}\epsilon \eta^3{+}\Big[\frac{e_r((15\nu{-}29)\nu e_r^2{-}328\nu{-}160 )}{32a_r^2(e_r^2{-}1)}-\frac{e_r(S+S_*)^2}{2a_r^2(e_r^2-1)}\epsilon^2\Big]\eta^4$\\
$e_t$&$e_r-\frac{e_r(3\nu-8)}{2a_r}\eta^2-\frac{e_r(2S+S_*)}{a_r^\frac{3}{2}\sqrt{e_r^2-1}}\epsilon \eta^3+\Big[\frac{e_r(-96+11\nu-15\nu^2)+e_r^3(128-67\nu+15\nu^2)}{8a_r^2(e_r^2-1)}+\frac{e_r(S+S_*)^2}{2a_r^2(e_r^2-1)}\epsilon^2\Big]\eta^4$\\
$f_t$&$-\frac{3(2\nu-5)}{2a_r^2\sqrt{e_r^2-1}}\eta^4$\\
$g_t$&$-\frac{e_r(\nu-15)\nu}{8a_r^2\sqrt{e_r^2-1}}\eta^4$\\
$f_\phi$&$-\frac{e_r^2(3\nu^2-19\nu-1)}{8a_r^2(e_r^2-1)^2}\eta^4$\\
$g_\phi$&$\frac{e_r^3\nu(1-3\nu)}{32a_r^2(e_r^2-1)^2}\eta^4$\\
$K$&$1{+}\frac{3}{a_r(e_r^2-1)}\eta^2{-}\frac{4S{+}3S_*}{a_r^\frac{3}{2}(e_r^2{-}1)^\frac{3}{2}}\epsilon\eta^3{+}\Big[\frac{42{-}24\nu{+}e_r^2(6\nu{-}9)}{4a_r^2(e_r^2{-}1)^2}+\frac{3(S+S_*)^2}{2a_r^2(e_r^2-1)^2}\epsilon^2\Big]\eta^4$\\
$\bar{n}$&$\frac{1}{a_r^\frac{3}{2}}-\frac{\nu-9}{2a_r^\frac{5}{2}}\eta^2-\frac{3(2S+S_*)}{2a_r^3\sqrt{e_r^2-1}}\epsilon\eta^3+\Big[\frac{3(-33-\frac{59\nu}{3}-\nu^2+e_r^2\left(49-\frac{25\nu}{3}+\nu^2\right))}{8a_r^\frac{7}{2}(e_r^2-1)}+\frac{3(S+S_*)^2}{4a_r^{7/2}(e_r^2-1)}\epsilon^2\Big]\eta^4$\\
 \end{tabular}
\end{ruledtabular}
\end{table*}
The conserved energy and angular momentum evaluated on the orbit give
\bea
\frac{\bar E}{\nu}&=& {\mathcal E}^{\rm orb}+\epsilon {\mathcal E}^{S}+ \epsilon^2{\mathcal E}^{SS}\,,\nonumber\\
\frac{j}{\nu}&=& {\mathcal J}^{\rm orb}+\epsilon {\mathcal J}^{S}+\epsilon^2 {\mathcal J}^{SS}\,,
\eea
with the various ingredients summarized in Table \ref{EandJ} (in units of $G=c=1$ and $M=1$, $L=\nu l$, $J=\nu j$, $E=\nu \bar{E}$).
%
%
\begin{table*}
\caption{\label{EandJ} Conserved energy, angular momentum as well as the orbital parameters  $a_r(\bar E,j)$ and $e_r(\bar E,j)$ in the aligned spin case, decomposed according to Eq. \eqref{deco_ar_er}.  
}
\begin{ruledtabular}
\begin{tabular}{ll}
$\mathcal{E}^{\rm orb}$ &
$\frac{1}{2a_r}-\frac{\nu-7}{8a_r^2}\eta^2+
\frac{7-49\nu-\nu^2+e_r^2(25-7\nu+\nu^2)}
{16a_r^3(e_r^2-1)}\eta^4$
\\

$\mathcal{E}^{S}$ &
$-\frac{(2S+S_*)}
{2a_r^{5/2}\sqrt{e_r^2-1}}\eta^3$
\\

$\mathcal{E}^{SS}$ &
$\frac{(S+S_*)^2}{4a_r^3(e_r^2-1)}
\eta^4$
\\
\hline
$\mathcal{J}^{\rm orb}$ &
$\sqrt{a_r}\sqrt{e_r^2-1}
-\frac{e_r^2(\nu-2)-4}
{2\sqrt{a_r}\sqrt{e_r^2-1}}\eta^2
+\frac{20-36\nu+e_r^4(12-11\nu+3\nu^2)
-e_r^2(4+53\nu+4\nu^2)}
{8a_r^{3/2}(e_r^2-1)^{3/2}}\eta^4$
\\

$\mathcal{J}^{S}$ &
$ S$
\\

$\mathcal{J}^{SS}$ &
$\frac{(3+e_r^2)(S+S_*)^2}
{4a_r^{3/2}(e_r^2-1)^{3/2}}
\eta^4$
\\
\hline
$a_r^{\rm orb}$ & $\frac{1}{2\bar{E}}+\frac{1}{4}(7-\nu)\eta^2+\Big[\frac{1}{8} \left(\nu ^2+1\right)
   \bar{E}+\frac{4-7 \nu }{L^2}\Big]\eta^4
$\\
$a_r^{S}$ & $-\frac{S_*+2 S}{L}\eta^3$\\
$a_r^{SS}$ & $\frac{(S+S_*)^2}{2L^2}\eta^4$\\
\hline
$e_r^{\rm orb}$ & $\sqrt{2 \bar E L^2+1}+\frac{\bar{E} \left(5 L^2 (\nu
   -3) \bar{E}+2 (\nu
   -6)\right)}{2 \sqrt{2 L^2
   \bar{E}+1}}\eta^2+\frac{\bar{E} \left(L^6 (7 (\nu
   -30) \nu +415) \bar{E}^3+4 L^4
   (\nu  (\nu +148)+50)
   \bar{E}^2+8 L^2 (99 \nu -35)
   \bar{E}+32 (7 \nu
   -4)\right)}{8 L^2 \left(2 L^2
   \bar{E}+1\right)^{3/2}}\eta^4$\\
$e_r^{S}$ & $-\frac{2 L \bar{E} S}{\sqrt{2
   L^2 \bar{E}+1}}+\frac{2 \bar{E} \left(L^2 \bar{E}
   \left(S_*+4S\right)+2
   S_*+4 S\right)}{L
   \sqrt{2 L^2 \bar{E}+1}}\eta^3$\\
$e_r^{SS}$ &
$-\frac{2(S+S_*)^2\bar{E}(1+L^2\bar{E})}{L^2\sqrt{2L^2\bar{E}+1}}\eta^4$
\\
\end{tabular}
\end{ruledtabular}
\end{table*}

As a by-product from the orbit one immediately gets the scattering angle. The $O(S^0)$ part agrees with existing literature. At $O(S^1)$ we find

\bea
\chi^{s^1}&=&-\frac{4 S \left(b^2 p_{\infty }^4+2\right)}{b^4 p_{\infty }^5+b^2 p_{\infty
   }}-\frac{S_* \left(4 b^2 p_{\infty }^4+6\right)}{b^4 p_{\infty }^5+b^2
   p_{\infty }}\nonumber\\
&+&\left(-\frac{16 S}{b^3 p_{\infty }^3}-\frac{12 S_*}{b^3
   p_{\infty }^3}\right)At\,,
\eea
with 
\beq
At={\rm arctan} \left(\frac{\sqrt{\sqrt{b^2 p_{\infty
   }^4+1}+1}}{\sqrt{\sqrt{b^2 p_{\infty }^4+1}-1}}\right)\,,
\eeq
i.e., in expanded form
\bea
\chi^{s^1}&=& \frac{2 \left(9 S+5 S_*\right)}{7 b^{10} p_{\infty }^{17}}-\frac{2 \left(7
   S+4 S_*\right)}{5 b^8 p_{\infty }^{13}}+\frac{\frac{10 S}{3}+2 S_*}{b^6
   p_{\infty }^9}\nonumber\\
&+&\frac{-6 S-4 S_*}{b^4 p_{\infty }^5}-\frac{\pi  \left(4
   S+3 S_*\right)}{2 b^3 p_{\infty }^3}-\frac{2 \left(S+S_*\right)}{b^2
   p_{\infty }}\nonumber\\
&+& O\left(\frac{1}{b^{11}}\right)\,,
\eea
whereas at $O(S^2)$ we find
\bea
\chi^{s^2}&=&\left(\frac{6 S^2}{b^4 p_{\infty }^4}+\frac{12 S_* S}{b^4 p_{\infty
   }^4}+\frac{6 S_*^2}{b^4 p_{\infty }^4}\right) At\nonumber\\
&+&\frac{S^2 \left(2 b^2 p_{\infty
   }^4+3\right)}{b^5 p_{\infty }^6+b^3 p_{\infty }^2}+\frac{S_* S \left(4
   b^2 p_{\infty }^4+6\right)}{b^5 p_{\infty }^6+b^3 p_{\infty
   }^2}\nonumber\\
&+&\frac{S_*^2 \left(2 b^2 p_{\infty }^4+3\right)}{b^5 p_{\infty
   }^6+b^3 p_{\infty }^2}\,,
\eea
that is, in expanded form
\bea
\chi^{s^2}&=& \left(S+S_*\right){}^2\left(\frac{4 }{5 b^9 p_{\infty
   }^{14}}-\frac{1}{b^7 p_{\infty }^{10}}+\frac{2}{b^5 p_{\infty }^6}\right.\nonumber\\
&+&\left. \frac{3 \pi}{4 b^4 p_{\infty
   }^4}+\frac{1}{b^3 p_{\infty }^2}\right)+O\left(\frac{1}{b^{10}}\right)\,.
\eea
These expressions suggest the   presence  of the natural, dimensionless variable
\beq
\zeta= b p_\infty^2\sim  \frac{c^2 b}{GM}\frac{p_\infty^2}{c^2}\,.
\eeq
For example,
\bea
\chi^{s^2}&=& \left(S+S_*\right){}^2p_\infty^4 f(\zeta)\,,
\eea
with
\bea
f(\zeta)= \frac{4 }{5 \zeta^9}-\frac{1}{\zeta^7}+\frac{2}{\zeta^5}+\frac{3 \pi}{4 \zeta^4  }+\frac{1}{\zeta^3}+O\left(\frac{1}{\zeta^{10}}\right)\,.
\eea

Within a different approach, Ref. \cite{Bini:2017wfr} was suggesting the use of the following dimensionless quantity 
\beq
\alpha =\frac{c}{p_\infty} \frac{Gm_1m_2}{c J}\,,
\eeq
to be kept as constant when expanding in large angular momentum (both in the spinless and the spinning case). Following this prescription, a direct comparison at level of final expressions is not possible and one should go back to the initial derivation details.

\section{Radiative losses}

Let us display the spin-corrected energy, angular momentum and linear momentum fluxes, which use the general expressions
\bea
\label{En_eq}
{\mathcal F}_E(t_r) &=& \frac{G}{c^5}\left\{
\frac15 U^{(1)}_{ij}U^{(1)}_{ij}\right.\nonumber\\
&+& \eta^2 \left[\frac{1}{189}U^{(1)}_{ijk}U^{(1)}_{ijk}+\frac{16}{45}V^{(1)}_{ij}V^{(1)}_{ij}\right]\nonumber\\
&+& \eta^4 \left[ \frac{1}{9072}U^{(1)}_{ijkm}U^{(1)}_{ijkm}+\frac{1}{84}V^{(1)}_{ijk}V^{(1)}_{ijk} \right]\nonumber\\
&+& \left. O(\eta^6)\right\}\,,
\eea
\bea
\label{J_eq}
{\mathcal F}_{J_i}(t_r) &=& \frac{G}{c^5}\epsilon_{iab}\left\{
\frac25 U_{aj}U^{(1)}_{bj}\right.\nonumber\\
&+& \eta^2 \left[\frac{1}{63}U_{ajk}U^{(1)}_{bjk}+\frac{32}{45}V_{aj}V^{(1)}_{bj}\right]\nonumber\\
&+& \eta^4 \left[ \frac{1}{2268}U_{ajkl}U^{(1)}_{bjkl}+\frac{1}{28}V_{ajk}V^{(1)}_{bjk} \right]\nonumber\\
&+& \left. O(\eta^6)\right\}\,,
\eea
and
\bea
\label{Pi_eq}
{\mathcal F}_{P_i}(t_r) &=& \frac{G}{c^7}\left\{
\frac2{63} U^{(1)}_{ijk}U^{(1)}_{jk}+\frac{16}{45}\epsilon_{ijk}U_{jl}^{(1)} V_{kl}^{(1)}\right.\nonumber\\
&+& \eta^2 \left[\frac{1}{1134}U^{(1)}_{ijkl}U^{(1)}_{jkl}+\frac{1}{126}\epsilon_{ijk}U^{(1)}_{jab}V^{(1)}_{kab}\right.\nonumber\\
&+&\left.\frac{4}{63}V^{(1)}_{ijk}V^{(1)}_{jk}\right]\nonumber\\  
&+& \eta^4 \left[ \frac{1}{59400}U^{(1)}_{ijklm}U^{(1)}_{jklm}+\frac{2}{14175}\epsilon_{ijk}U^{(1)}_{jabc}V^{(1)}_{kabc}\right.\nonumber\\
&+&\left.\left. \frac{2}{945}V^{(1)}_{ijkl}V^{(1)}_{jkl} \right]+ O(\eta^6)\right\}\,.
\eea
Even if the general relations \eqref{En_eq}, \eqref{J_eq}, \eqref{Pi_eq} involving radiative multipoles $(U_L,V_L)$ are the same in the spinless case, when one re-expresses the radiative multipoles in terms of source multipoles $(I_L,J_L)$,
\beq
U_L=I_L^{(l)}+O(\eta^3)\,,\qquad V_L=J_L^{(l)}+O(\eta^3)\,,
\eeq
(with $L=i_1\ldots i_l$ a multi-index notation)
there exist explicit spin-orbit and spin-spin corrections as shown in Tab. \ref{Spin_corr_multipoles} and \ref{Spin_corr_multipoles2}. Obviously, when evaluating these expressions along a spin-corrected trajectory, spin terms arise also from the spinless part.

\subsection{Radiated energy}
The flux of radiated energy is given by
\beq
\label{dE_dt_eq}
\frac{dE}{dt}={\mathcal F}_E^{s^0}+{\mathcal F}_E^{s^1}+{\mathcal F}_E^{s^2}\,,
\eeq
where ${\mathcal F}_E^{s^0}$ can be found in Ref. \cite{Arun:2007sg}, Eq. (5.2a)-(5.2e), ${\mathcal F}_E^{s^1}$ can be found in Ref. \cite{Bini:2023mdz}, Eq. (16) and
\bea
{\mathcal F}_E^{s^2}&=& \frac{\nu^2 \eta^4}{(1-4\nu)r^4}\left[ \left( \frac{32}{5}\nu \frac{\dot r^2}{ r^2}   + 
 \frac{98}{5}\dot \phi^2   - 
 \frac{384}{5} \nu \dot \phi^2  \right)(S^2+S_*^2)\right.\nonumber\\ 
&+&\left. \left(  -\frac{32}{5}(1-2\nu )\frac{\dot r^2}{r^2}   + \left(\frac{188}{5} -  \frac{768}{5}\nu \right)\dot \phi^2\right)SS_*\right]\,.
\eea

The previous equation can be integrated exactly leading to $\Delta E_{\rm rad}=\Delta E_{\rm rad}^{s^0}+\Delta E_{\rm rad}^{s^1}+\Delta E_{\rm rad}^{s^2}$. The complete (exact) expression $\Delta E_{\rm rad}$ is given in the associated ancillary file.
In a large $b$ expansion limit we find
\bea
\Delta E_{\rm rad}^{s^2}=\Delta E_{\rm rad}^{s^2, \slashed {\pi}}+\pi \Delta E_{\rm rad}^{s^2, \pi}\,,
\eea
with
\begin{widetext}
\bea
\Delta E_{\rm rad}^{s^2, {\slashed \pi}}&=&  \left[\left(\frac{127744 \nu }{225}-\frac{32896}{225}\right)(S^2 +S_*^2)+\left(\frac{255488 \nu }{225}-\frac{61952}{225}\right) S_* S\right]\frac{1}{b^6 p_\infty}\nonumber\\
&+& \left[\left(\frac{73472 \nu }{15}-\frac{18752}{15}\right)(S^2+ S_*^2)+\left(\frac{146944 \nu }{15}-\frac{35968}{15}\right) S_* S \right] \frac{1}{b^8 p_\infty^5}\nonumber\\
&+& \left[ \left(\frac{675328 \nu }{75}-\frac{34304}{15}\right)(S^2+ S_*^2)+\left(\frac{1350656 \nu }{75}-\frac{332288}{75}\right) S_* S \right]\frac{1}{b^{10} p_\infty^9}\nonumber\\
&+&  \left[\left(\frac{958976 \nu }{315}-\frac{34688}{45}\right) (S^2+S_*^2)+\left(\frac{1917952 \nu }{315}-\frac{473344}{315}\right) S_* S\right]\frac{1}{b^{12} p_\infty^{13}}\nonumber\\
&{+}&O(\frac{1}{b^{14}})\,,
\eea
and
\bea
\label{noi_s2}
\Delta E_{\rm rad}^{s^2, \pi}&=&\left[ \left(\frac{118 \nu }{5}-\frac{49}{8}\right)( S^2+S_*^2)+\left(\frac{236 \nu }{5}-\frac{227}{20}\right) S_* S \right] \frac{p_\infty}{b^5}\nonumber\\
&+& \left[  \left(668 \nu -\frac{1369}{8}\right) (S^2+S_*^2)+\left(1336 \nu -\frac{1303}{4}\right) S_* S \right] \frac{1}{p_\infty^3 b^7}\nonumber\\
&+& \left[ \left(\frac{7546 \nu }{3}-\frac{15365}{24}\right) (S^2+S_*^2)+\left(\frac{15092 \nu }{3}-\frac{14819}{12}\right) S_* S \right] \frac{1}{p_\infty^7 b^9}\nonumber\\
&+& \left[ \left(\frac{11088 \nu }{5}-\frac{22491}{40}\right) (S^2+S_*^2)+\left(\frac{22176 \nu }{5}-\frac{21861}{20}\right) S_* S \right] \frac{1}{p_\infty^{11} b^{11}}\nonumber\\
&+& O\left(\frac{1}{b^{13}}\right)\,.
\eea

$\Delta E_{\rm rad}^{s^0}$ can be found in Ref. \cite{Bini:2021gat} Eqs. (C11)-(C13) and $\Delta E_{\rm rad}^{s^1}$ can be found in Ref. \cite{Bini:2023mdz}, Eq. (25).
\subsection{Radiated angular momentum}

Similarly, the flux of angular momentum is
\bea
\frac{dJ}{dt}
={\mathcal F}_J^{s^0}+{\mathcal F}_J^{s^1}+{\mathcal F}_J^{s^2}\,,
\eea
where ${\mathcal F}_J^{s^0}$ can be found in Ref. \cite{Arun:2009mc},  Eqs. (3.4a) -(3.4c), ${\mathcal F}_J^{s^1}$ can be found in  \cite{Bini:2023mdz}, Eq. (16) and

\bea
{\mathcal F}_J^{s^2}&=&\dot \phi \left\{ (S^2+S_*^2) \left[\frac{1}{r^4}\left(\frac{48}{5}+\frac{2}{5 (1-4 \nu)}\right) -\frac{36 \dot r^2 }{5 r^3}+\frac{24\dot \phi^2}{5 r}\right]\right.\nonumber\\
&+&S_* S \left[\frac{1}{r^4}\left(-\frac{4}{5 (1-4 \nu)
   }+\frac{96}{5}\right) -\frac{72 \dot r^2 }{5 r^3}+\frac{48\dot \phi^2}{5 r}\right]\,. 
\eea
The previous equation can be integrated exactly leading to $\Delta J_{\rm rad}=\Delta J_{\rm rad}^{s^0}+\Delta J_{\rm rad}^{s^1}+\Delta J_{\rm rad}^{s^2}$. The complete (exact) expression $\Delta J_{\rm rad}$ is given in the associated ancillary file.
In a large $b$ expansion limit we find
\bea
\Delta J_{\rm rad}^{s^2}=\Delta J_{\rm rad}^{s^2, \slashed {\pi}}+\pi \Delta J_{\rm rad}^{s^2, \pi}\,,
\eea
with
\bea
\Delta J_{\rm rad}^{s^2, {\slashed \pi}}&=& \frac{\nu^2}{(1-4\nu)}\left\{\left[\left(\frac{16}{5}-\frac{64 \nu }{5}\right) (S^2+S_*^2)+\left(\frac{32}{5}-\frac{128 \nu }{5}\right) S_* S  \right]\frac{p_\infty^2}{b^3}\right.\nonumber\\
&+& \left[\left(\frac{832}{5}-\frac{9856 \nu }{15}\right) (S^2+S_*^2)+\left(\frac{4864}{15}-\frac{19712 \nu }{15}\right) S_* S  \right]\frac{1}{p_\infty^2b^5} \nonumber\\
&+&\left. \left[ \left(\frac{2768}{5}-\frac{10944 \nu }{5}\right) (S^2+S_*^2)+\left(\frac{5408}{5}-\frac{21888 \nu }{5}\right) S_* S \right]\frac{1}{p_\infty^6b^7}
\right\} \nonumber\\
&+&\Big[\Big(256-\frac{25344\nu}{25}\Big)(S^2+S_*^2)+\Big(\frac{12544}{25}-\frac{50688\nu}{25}\Big)SS_*\Big]\frac{1}{b^9p_\infty^{10}} \nn\\
&+&\Big[\Big(-\frac{2448}{35}+\frac{832\nu}{3}\Big)(S^2+S_*^2)+\Big(-\frac{14432}{105}+\frac{1664\nu}{3}\Big)SS_*\Big]\frac{1}{b^{11}p_\infty^{14}} \nn\\
&+&O\left(\frac{1}{b^{13}} \right)\,,
\eea
and
\bea
\Delta J_{\rm rad}^{s^2, \pi}&=&\frac{\nu^2}{(1-4\nu)}\Big\{ 
\left[ \left(\frac{51}{4}-\frac{252 \nu }{5}\right) (S^2+S_*^2)+\left(\frac{249}{10}-\frac{504 \nu }{5}\right) S_* S \right] \frac{1}{b^4} \nonumber\\
&+&  \left[\left(\frac{243}{2}-480 \nu \right) (S^2+S_*^2)+(237-960 \nu ) S_* S  \right] \frac{1}{p_\infty^4 b^6}\nonumber\\
&+&\Big[\Big(\frac{651}{4}-644\nu\Big)(S^2+S_*^2)+\Big(\frac{637}{2}-1288\nu\Big)SS_* \Big]\frac{1}{b^8p_\infty^8}\nn\\
&+&  O\left(\frac{1}{b^{14}}\right)\Big\}\,.
\eea

$\Delta J_{\rm rad}^{s^0}$  can be found in Ref. \cite{Bini:2021gat},  Eqs. (E8) -(E10), $\Delta J^{s^1}$ can be found in  \cite{Bini:2023mdz}, Eq. (26).

\subsection{Radiated linear momentum}

The flux of radiated linear momentum is given by
\beq
\label{dP_dt_eq}
\frac{dP_i}{dt}= {\mathcal F}_{P_i}^{s^0}+{\mathcal F}_{P_i}^{s^1}+{\mathcal F}_{P_i}^{s^2}\,,
\eeq
with (the only relevant)
\bea
{\mathcal F}_{P_y}^{s^2}&=& \frac{\nu^2}{\sqrt{1-4\nu}}\left\{\frac{\dot \phi^3 \cos (\phi)}{r^3}\left[ \frac{8}{35} (208 \nu -41) S^2
      +\frac{64}{105} (78 \nu -25)
   S_*^2    +\frac{8}{105} (1248
   \nu -301) S S_*  
    \right] \right.\nonumber\\
&+&\frac{\dot r \dot \phi^2 \sin (\phi )}{r^4}\left[ -\frac{4}{35} (232 \nu -185)
    S^2  
   -\frac{4}{105}
   (696 \nu +193)   S_*^2
   -\frac{8}{105} (696 \nu
   -167)  S S_*   \right]\nonumber\\
&+&\frac{\dot{r}^2  \dot{\phi }\cos (\phi )}{r^5}\left[ -\frac{8}{35} (16 \nu -5)
    S^2  
    -\frac{8}{105}
   (48 \nu -1)  S_*^2
    -\frac{256}{105} (3 \nu -1)
    S S_*  
     \right]\nonumber\\
&+&\frac{1}{r^6}\left[ -\frac{8}{35} S^2 \left((1-40
   \nu ) \dot{\phi } \cos (\phi
   )+2 (2 \nu -1) \dot{r}^3 \sin
   (\phi )\right)-\frac{4}{105}
   S_* S \left(6 (23-80 \nu )
   \dot{\phi } \cos (\phi )+4
   (12 \nu -1) \dot{r}^3 \sin
   (\phi )\right)\right.\nonumber\\
&-&\left.\frac{32}{105}
   S_*^2 \left(6 (2-5 \nu )
   \dot{\phi } \cos (\phi )+(3
   \nu -1) \dot{r}^3 \sin (\phi
   )\right) \right]\nonumber\\
&+&\left.\frac{\dot r \sin (\phi)}{r^7}\left[ -\frac{16}{35} (2 \nu -3)
     S^2  -\frac{16}{105} (12 \nu +5)
    S_* S  -\frac{32}{105} (3 \nu -1)
    S_*^2  \right]\right\}\,,
\eea
where we observe that, differently from the energy and angular momentum losses, the   combinations $S^2+S_*^2$ and $S S_*$ do not play any special role.
The previous equation can be integrated exactly leading to $\Delta P_{i\,\rm rad}=\Delta P_{i\,\rm rad}^{s^0}+\Delta P_{i\,\rm rad}^{s^1}+\Delta P_{i\,\rm rad}^{s^2}$
with
\beq
\Delta P_{x\,\rm rad}= 0\,,
\eeq
(only its instantaneous part vanishes up to the PN level of interest here, as shown in Ref. \cite{Bini:2021gat})
and the complete (exact) expression $\Delta P_{y\,\rm rad}$ is given in the associated ancillary file.
In a large $b$ expansion limit we find
\bea
\Delta P_{y\,\rm rad}^{s^2}=\Delta P_{y\,\rm rad}^{s^2, \slashed {\pi}}+\pi \Delta P_{y\,\rm rad}^{s^2, \pi}\,,
\eea
with   
\bea
\Delta P_{y\,\rm rad}^{s^2, {\slashed \pi}}&{=}&\frac{\nu^2}{\sqrt{1{-}4\nu}}\Big\{\Big[S_*^2\Big({-}\frac{66368}{525}{+}\frac{522496\nu}{1575}\Big){+}S^2\Big({-}\frac{21824}{525}{+}\frac{522496\nu}{1575}\Big){+}SS_*\Big({-}\frac{51584}{315}{+}\frac{1044992\nu}{1575}\Big)\Big]\frac{1}{b^6}\nn\\
&{+}&\Big[S_*^2\Big({-}\frac{2148064}{1575}{+}\frac{243584\nu}{63}\Big){+}S^2\Big({-}\frac{199712}{315}{+}\frac{243584\nu}{63}\Big){+}SS_*\Big({-}\frac{980992}{525}{+}\frac{487168\nu}{63}\Big)\Big]\frac{1}{b^8p_\infty^4}\nn\\
&{+}&\Big[S_*^2\Big({-}\frac{603464}{175}{+}\frac{16126816\nu}{1575}\Big){+}S_*^2\Big({-}\frac{2911912}{1575}{+}\frac{16126816\nu}{1575}\Big){+}SS_*\Big({-}\frac{2594576}{525}{+}\frac{32253632\nu}{1575}\Big)\Big]\frac{1}{b^{10}p_\infty^8}\nn\\
&{+}&\Big[S_*^2\Big({-}\frac{3725308}{1575}{+}\frac{11450416\nu}{1575}\Big){+}S^2\Big({-}\frac{2188868}{1575}{+}\frac{11450416\nu}{1575}\Big){+}SS_*\Big({-}\frac{1107248}{315}{+}\frac{22900832\nu}{1575}\Big)\Big]\frac{1}{b^{12}p_\infty^{12}}\nn\\
&+&O\Big(\frac{1}{b^{14}}\Big)\,,
\eea
and
\bea
\Delta P_{y\,\rm rad}^{s^2, \pi}&=& \frac{\nu^2}{\sqrt{1-4\nu}}\Big\{\Big[S_*^2\Big(-\frac{155}{32}+\frac{59\nu}{5}\Big)+S^2\Big(-\frac{141}{160}+\frac{59\nu}{5}\Big)+SS_*\Big(-\frac{243}{40}+\frac{118\nu}{5}\Big)\Big]\frac{p_\infty^2}{b^5}\nn\\
&+&\Big[S_*^2\Big(-\frac{52793}{320}+\frac{6794}{15}\Big)+S^2\Big(-\frac{65129}{960}+\frac{6794\nu}{15}\Big)+SS_*\Big(-\frac{17609}{80}+\frac{13588\nu}{15}\Big)\Big]\frac{1}{p_\infty b^7}\nn\\
&{+}&\Big[S_*^2\Big({-}\frac{3097487}{3840}{+}\frac{93807\nu}{40}\Big){+}S^2\Big({-}\frac{312449}{768}{+}\frac{93807\nu}{40}\Big){+}SS_*\Big({-}\frac{72429}{64}{+}\frac{93807\nu}{20}\Big)\Big]\frac{1}{b^9p_\infty^6}\nn\\
&{+}&\Big[S_*^2\Big({-}\frac{8383111}{7680}{+}\frac{396119\nu}{120}\Big){+}S^2\Big({-}\frac{4722493}{7680}{+}\frac{396119\nu}{120}\Big){+}SS_*\Big({-}\frac{1020501}{640}{+}\frac{396119\nu}{60}\Big)\Big]\frac{1}{b^{11}p_\infty^{10}}\nn\\
&+&O\Big(\frac{1}{b^{13}}\Big)\,.
\eea

\end{widetext}

\section{Radial infall in the aligned spins case}

Let us study the case ${\bf S}=(0,0,0)$ which correspond to ${\bf S}_1=(0,0,s)$ and ${\bf S}_2=(0,0,-s)$. As a consequence we have ${\bf \Delta}=(0,0,-s/\nu)=\frac{M}{\delta m}{\mathbf S}_*=\frac{1}{\sqrt{1-4\nu}}{\mathbf S}_*$ so that ${\mathbf S}_*=(0,0,-\sqrt{1-4\nu}/\nu)s$. The total acceleration up to the 2PN accuracy is given in Eq. \eqref{rel_acc}. 

As a consequence of this spin choices the motion remains purely radial (at the present PN accuracy). A direct integration gives the following (conservative plus radiation-reacted) orbit (using as a parameter the dimensionless time $T=t/M$)
\bea\label{orbRF}
\frac{r(t)}{M}&=&\frac{3^\frac{2}{3}T^\frac{2}{3}}{2^\frac{1}{3}}+\frac{5}{2}(\nu-2)\eta^2\nonumber\\
&+&\frac{\eta^4}{2\cdot 6^\frac{2}{3}T^\frac{2}{3}}(48-19\nu+5\nu^2)\nonumber\\  
&-& \hat{s}^2\eta^4\frac{1-4\nu}{6^\frac{2}{3}T^\frac{2}{3}\nu}\,,
\eea
and where we introduced the dimensionless spin variable 
\beq
\hat s=\frac{s}{M^2}\,,
\eeq
which, with an abuse of notation, we still denote as $s$ when working in units of $M$.

Note that linear-in-spin correction to the orbits in this special case of radial motion and constant spin orthogonal to the orbital plane cancel out.
Let's emphasize that in the previous expression both $s$ and $r$ are dimensionfull.

\subsection{Radial infall and aligned spins}

Specializing the above expressions for radial orbits and for aligned spins orthogonal to the orbital plane leads to the following expressions where all multipole moments are proportional to constant STF tensors,
\bea
I_{ij}^{S}&=&\frac{2^\frac{4}{3}s T^\frac{1}{3}\sqrt{1-4\nu}}{3^\frac{2}{3}}\eta^3\delta^{I_2, S}_{ij}\,,
\nonumber\\
I_{ij}^{SS}&=&\frac{s^2\eta^4}{3\nu}\delta^{I_2, SS}_{ij}\,, \nonumber\\
J_{ij}^{S}&=&\Big[\frac{3^\frac{5}{3}s T^\frac{2}{3}}{2^\frac{7}{3}}\eta-\Big(\frac{195}{56}+\frac{33\nu}{28}\Big)s\eta^3\Big]
\delta^{J_2, S}_{ij}\,,
\nonumber\\
J_{ij}^{SS}&=&\frac{s^2\eta^4\sqrt{1-4\nu}}{6^\frac{1}{3}T^\frac{1}{3}\nu} \delta^{J_2, SS}_{ij}\,,
\nonumber\\
I_{ijk}^{S}&=&\frac{6}{5} s T \eta^3(5\nu-1)\delta^{I_3,S}_{ijk}\,,
\nonumber\\
I_{ijk}^{SS}&=&-\frac{3^\frac{5}{3}s^2T^\frac{2}{3}\eta^4\sqrt{1-4\nu}}{5\cdot 2^\frac{1}{3}\nu} \delta^{I_3, SS}_{ijk}\,,
\nonumber\\
J_{ijk}^{S}&=&-\frac{4}{5}6^\frac{1}{3}s T^\frac{4}{3}\eta\sqrt{1-4\nu}
\delta^{J_3, S}_{ijk}\,,
\eea
where $T=\frac{t}{M}$ is a convenient dimensionless time variable 
\bea
\delta^{I_2, S}_{ij}&=& \left(
\begin{array}{ccc}
 0 & 1 & 0 \\
 1 & 0 & 0 \\
 0 & 0 & 0 \\
\end{array}
\right) \,,\qquad
\delta^{I_2, SS}_{ij}= \left(
\begin{array}{ccc}
 1 & 0 & 0 \\
 0 & 1 & 0 \\
 0 & 0 & -2 \\
\end{array}
\right) \,,\nonumber\\
\delta^{J_2, S}_{ij}&=&
\left(
\begin{array}{ccc}
 0 & 0 & 1 \\
 0 & 0 & 0 \\
 1 & 0 & 0 \\
\end{array}
\right)\,,\qquad
\delta^{J_2, SS}_{ij}= \left(
\begin{array}{ccc}
 0 & 0 & 0 \\
 0 & 0 & 1 \\
 0 & 1 & 0 \\
\end{array}
\right)
\eea
Moreover
\bea
\delta^{I_3, S}_{ijk}&=&\left(
\begin{array}{ccc}
 \{0,1,0\} & \{1,0,0\} &
   \{0,0,0\} \\
 \{1,0,0\} &
   \left\{0,-\frac{3}{4},0\right
   \} &
   \left\{0,0,-\frac{1}{4}\right
   \} \\
 \{0,0,0\} &
   \left\{0,0,-\frac{1}{4}\right
   \} &
   \left\{0,-\frac{1}{4},0\right
   \} \\
\end{array}
\right) \,,\nonumber\\
\delta^{I_3, SS}_{ijk}&=&\left(
\begin{array}{ccc}
 \{1,0,0\} &
   \left\{0,\frac{1}{3},0\right\} &
   \left\{0,0,-\frac{4}{3}\right
   \} \\
 \left\{0,\frac{1}{3},0\right\}
   &
   \left\{\frac{1}{3},0,0\right\} & \{0,0,0\} \\
 \left\{0,0,-\frac{4}{3}\right\}
   & \{0,0,0\} &
   \left\{-\frac{4}{3},0,0\right\} \\
\end{array}
\right)\,.
\eea
Finally,
\beq
\delta^{J_3, S}_{ijk}=
 \left(
\begin{array}{ccc}
 \{0,0,1\} & \{0,0,0\} &
   \{1,0,0\} \\
 \{0,0,0\} &
   \left\{0,0,-\frac{1}{4}\right
   \} &
   \left\{0,-\frac{1}{4},0\right
   \} \\
 \{1,0,0\} &
   \left\{0,-\frac{1}{4},0\right
   \} &
   \left\{0,0,-\frac{3}{4}\right
   \} \\
\end{array}
\right)\,.
\eeq

In order to explain the matrix notation introduced above  let us consider as an example the electric octupolar multipole moment $I_{ijk}^S$. It is proportional to a constant   $3\times3$ matrix $\delta_{ijk}^{I_3,S}$  whose elements are not scalars but three $1\times 3$ vectors. Our notation is the following:  the first two indices represent the position (row/column) of the array inside the $3\times3$ matrix and the third index the position of the element inside the array.
The present notation can be also adapted to a STF tensor with four indices: it will be represented in terms of a $3\times3$ matrix whose elements are again $3\times3$ matrices with the first two indices  representing the position of the building block matrix inside the whole tensor, while the last two indices depict the position inside the building block $3\times3$ matrix. For better clarity, besides this explicit matrix-like representation we list below all the independent components of each of these tensors.

More precisely, denoting generically by $T$ a STF tensor,  we recall  that: 1)  a STF two-indices tensor in $d=3$ has only 5 independent components (e.g., $T_{11}$, $T_{22}$, $T_{12}$, $T_{13}$, $T_{23}$); 2) a three-indices STF tensor in $d=3$ has only 7 independent components (e.g., $T_{111}$, $T_{112}$, $T_{113}$, $T_{122}$, $T_{123}$, $T_{222}$, $T_{223}$). 
Therefore, for the above constant tensors the independent components are given by
\bea
&& \delta_{111}^{I_3,S}=0,\quad \delta_{112}^{I_3,S}=1,\quad \delta_{113}^{I_3,S}=0,\quad \delta_{122}^{I_3,S}=0,\nonumber\\
&& \delta_{123}^{I_3,S}=0,\quad \delta_{222}^{I_3,S}=-\frac34,\quad \delta_{223}^{I_3,S}=0\,,
\eea 

\bea
&& \delta_{111}^{I_3,SS}=1,\quad \delta_{112}^{I_3,SS}=0,\quad \delta_{113}^{I_3,SS}=0,\quad \delta_{122}^{I_3,SS}=\frac13,\nonumber\\
&& \delta_{123}^{I_3,SS}=0,\quad \delta_{222}^{I_3,SS}=0,\quad \delta_{223}^{I_3,SS}=0\,,
\eea

\bea
&& \delta_{111}^{J_3,S}=0,\quad \delta_{112}^{J_3,S}=0,\quad \delta_{113}^{J_3,S}=1,\quad \delta_{122}^{J_3,S}=0,\nonumber\\
&& \delta_{123}^{J_3,S}=0,\quad \delta_{222}^{J_3,S}=0,\quad \delta_{223}^{J_3,S}=-\frac14\,,
\eea 

\bea
&& \delta_{111}^{J_3,SS}=0,\quad \delta_{112}^{J_3,SS}=0,\quad \delta_{113}^{J_3,SS}=0,\quad \delta_{122}^{J_3,SS}=0,\nonumber\\
&& \delta_{123}^{J_3,SS}=1,\quad \delta_{222}^{J_3,SS}=0,\quad \delta_{223}^{J_3,SS}=0\,,
\eea

The energy and linear momentum fluxes for the radial infalling trajectory appearing in \cite{Bini:2026ova,DiRusso:2026fqn} acquire spin corrections only at the second order in spin as we will show below. This is a peculiar characteristic of the radial infall, which, in a sense, is simply  explained by the vanishing of the angular momentum (and hence of any spin-orbit coupling) in this specific case and at the accuracy level considered here.

The spin-quadratic correction to the energy flux at infinity for radially infalling trajectory is
\beq
\frac{dE^{(\eta^4, s^2)}_{\rm rad}}{dt}=\frac{64\cdot 2^\frac{2}{3}}{405\cdot 3^\frac{2}{3}T^\frac{14}{3}}s^2\eta^4\,,
\eeq
whereas the $O(S^0)$ part can be found in Eqs. (5.10)-(5.11) of Ref. \cite{DiRusso:2026fqn}\,.

In the spinless case both the conserved and the radiated angular momentum  were identically vanishing. In the presence of spin, we have instead the only nonvanishing component along the $z$ axis given by
\beq
\frac{dJ_{z\,\rm rad}}{dt}=\frac{16\cdot 2^{2/3}}{135\cdot 3^{1/3}}\frac{\nu\sqrt{1-4\nu}}{T^{10/3}}s \epsilon \eta^3\,.
\eeq
Finally, for the linear momentum losses we find
\bea
\frac{dP_{x\,\rm rad}^{s^2}}{dt}&=&\nu^2 \sqrt{1-4\nu}\eta^7 \left[\frac{2^{2/3}\eta^2}{3^{1/3} T^{13/3}} \left(\frac{640  \nu }{15309
    }-\frac{2720 
   }{15309 
    } \right)\right.\nonumber\\
&+&\frac{\eta^4}{T^5} \left(\frac{128 s^2 \epsilon ^2}{5103
   \nu ^2  }-\frac{256 s^2
   \epsilon ^2}{3645 \nu
   }\right.\nonumber\\
&-& \left. \left.
 \frac{512 \nu ^2}{6237
    }+\frac{100352 \nu }{120285
    }-\frac{17543872}{7577955
    } \right)\right]\,,\nonumber\\
\frac{dP_{y\,\rm rad}^{s}}{dt}&=& s\epsilon \nu \eta^7 \left[ \eta  \frac{128}{1215 T^4}  \right.\nonumber\\ 
&+&\left.\frac{2^{1/3}\eta^3}{ 3^{2/3} T^{14/3}}\left( \frac{2816 }{2835\
   }-\frac{2048
    \nu }{8505\
    }\right)\right]\,.
\eea
As expected, increasing the PN accuracy the rising of radiative losses effects will imply that the motion will not remain radial.

\section{Validation checks and benchmarks for future computations}

Ref. \cite{Jakobsen:2021lvp} (see   Eq. (31), with $a_1=\frac{S_1}{m_1}$ and $a_2=\frac{S_2}{m_2}$ aligned with the $z$ axis $e_3$, and assuming $C_{E,1}=0$, i.e., no tidal deformations) has shown that in this case
\begin{widetext}
\bea
E_{\rm rad}^{\rm LO}&=& \frac{G^3m_1^2m_2^2\pi}{b^3}v \left[
\frac{37}{15}+v\frac{(65m_1+69m_2)a_1+(65m_2+69m_1)a_2}{10 b M}+\frac{3920 a_1a_2+72a_1^2+72 a_2^2}{320b^2}
+O(v^2)\right]\,.
\eea
Passing to $S$ and $S_*$ (and working in units of $M=1=G$)
\bea
E_{\rm rad}^{\rm LO}&=& \frac{\nu^2\pi}{b^3}v \left[
\frac{37}{15}+v\frac{(65m_1+69m_2)  \frac{S_1}{m_1}+(65m_2+69m_1) \frac{S_2}{m_2}}{10 b }+  \frac{3920 \frac{S_1S_2}{m_1m_2}+72\frac{S_1^2}{m_1^2}+72 \frac{S_2^2}{m_2^2}}{320b^2}
+O(v^2)\right]
\nonumber\\
&=& \frac{\nu^2\pi}{b^3}v \left[
\frac{37}{15}+\frac{v}{b}\left(\frac{13}{2}S+\frac{69}{10}S_* \right)  
+\frac{1}{(1-4\nu)b^2}\underbrace{\left[ \left( \frac{49}{8}-\frac{118}{5}\nu\right)(S+S_*)^2-\frac{9}{10}S S_* \right]}_{\Lambda}
+O(v^2)\right]\,,
\eea
where we recall the standard relation between $v$ and $p_\infty$ 
\beq
v=\frac{p_\infty}{\sqrt{1+p_\infty^2}}\,. 
\eeq
The $O(S^2)$ term $\Lambda$ in the equation above
can be cast in the form 
\bea
\Lambda= \left(\frac{118 \nu }{5}-\frac{49}{8}\right)( S^2+S_*^2)+\left(\frac{236 \nu }{5}-\frac{227}{20}\right) S_* S  \,,
\eea
and agrees with Eq. \eqref{noi_s2} above.
Note that Ref. \cite{Bini:2023mdz} at the spin-orbit level confirmed
\beq
E_{\rm rad}^{\rm LO}=\nu^2 \frac{p_\infty^2\pi}{b^4}\left(\frac{13}{2}S+\frac{69}{10}S_* \right)\,,
\eeq
namely, the agreement with Ref. \cite{Jakobsen:2021lvp}.

\end{widetext}

\section{Conclusions}
We have studied a spinning two-body system in the simplified situation of aligned spins and for motions along hyperboliclike orbits.
We have computed  all radiative losses  using the MPM formalism: energy, angular momentum and linear momentum at fractional 2PN accuracy and including spin  corrections up to $O(S^2)$.
Leading PM order results are checked against existing literature, whereas higher-order PM results (within the 2PN accuracy) are new with this work and will be useful for future checks of similar expressions obtained within different formalisms (e.g. amplitudes).
As a by-product of our general results we have analyzed the spinning situation which supports radial infall, showing that as soon as the PN accuracy increases there no way to stay radial in presence of spin, even in the conservative case.

Future works will address (always in a scattering situation) the inclusion of radiation-reaction effects, which come together with NLO spin orbit corrections at 2.5PN accuracy. We expect, in this case, the loss of this (quite simple) picture and the necessity of considering precession effects, together with a variation of all the involved constants appearing up to the 2PN level studied here.

In addition, it will be interesting to explore the radiative losses in presence of (higher order corrections in) spin within a Fourier space description (e.g., the angular dependent energy spectrum), as well as the full waveform. 
This will be the object of future studies.

\section*{Acknowledgments}
D.B. thanks G. Faye and Q. Henry as well as A. Geralico for informative discussions.
D.B. and G.D.R. acknowledge  membership to the Italian Gruppo Nazionale per
la Fisica Matematica (GNFM) of the Istituto Nazionale
di Alta Matematica (INDAM).

\end{document}